\documentclass[12pt]{article}
\usepackage{amscd}
\usepackage{verbatim}
\usepackage{amssymb}
\usepackage{amsmath}
\newcommand{\qed}{\hfill $\Box$ \medskip}

\begin{document}
\begin{center}
\vspace{24pt} { \large \bf Investigating Off-shell Stability of
Anti-de Sitter Space in String Theory
} \\
\vspace{30pt}
{\bf V Suneeta} \footnote{vardarajan@math.ualberta.ca}\\
\vspace{24pt} 
{\footnotesize
Dept of Mathematical and Statistical Sciences \\ and \\ The Applied Mathematics Institute,\\ University of Alberta,
Edmonton, AB, Canada T6G 2G1.}
\end{center}
\date{\today}
\bigskip

\begin{center}
{\bf Abstract}
\end{center}
\noindent We propose an investigation of stability of vacua in
string theory by studying their stability with respect to a
(suitable) world-sheet renormalization group (RG) flow. We prove
geometric stability of (Euclidean) anti-de Sitter (AdS) space (i.e.,
$\mathbf{H}^n$) with respect to the simplest RG flow in closed
string theory, the Ricci flow. AdS space is not a fixed point of
Ricci flow. We therefore choose an appropriate flow for which it is
a fixed point, prove a linear stability result for AdS space with
respect to this flow, and then show this implies its geometric
stability with respect to Ricci flow. The techniques used can be
generalized to RG flows involving other fields. We also discuss
tools from the mathematics of geometric flows that can be used to
study stability of string vacua. \setcounter{equation}{0}
\newpage

\section{Introduction}

 \setcounter{equation}{0}
It has been conjectured that off-shell processes in string theory
such as tachyon condensation are described qualitatively by
solutions to RG flows of world-sheet sigma models (see \cite{HMT}
for a review). This conjecture is non-trivial because it implies
that the RG flow parameter plays the role of a dynamical time. There
are several arguments suggesting this connection, at least close to
a fixed point of the RG flow --- a fixed point being an on-shell or
vacuum geometry in string theory. In the world-sheet sigma model for
the on-shell or fixed point geometry, it is easy to construct
generic operators that cause excitations which are both tachyonic
and as well, relevant perturbations resulting in RG flows. An
example where the tachyonic excitation is marginal appears in
\cite{HMT} (page 6) where one considers a target space $M =
M_{1}^{d} \times R^{9-d,1}$, such that the world-sheet sigma model
on $M$ is a $CFT$ (of the form $CFT_1 + CFT_2$ ). One can construct
composite operators in the sigma model on $M$ comprising of operator
$O_1$ from $CFT_1$ and $O_2$ from $CFT_2$. In \cite{HMT}, a
composite operator is chosen such that it is a marginal operator in
the full CFT, the resultant excitation is tachyonic, and $O_1$ is a
relevant perturbation of $CFT_1$ causing an RG flow starting from
$M_1$. There is therefore a link between tachyonic instability of
$M$ (which could lead to processes such as tachyon condensation) and
instability of $M_1$ with respect to an RG flow (due to the
perturbation caused by $O_1$).

Tachyon condensation can occur both as an on-shell process and
through off-shell interactions. One example of the change in
geometry caused due to tachyon condensation is the orbifold flow
$C/Z_{n} \rightarrow C$
--- this is studied as an on-shell process in \cite{APS} and as an
off-shell process in, for example, \cite{OZ}. There exists a
solution to an RG flow  that describes this change in geometry
qualitatively
--- this exact solution (to the {\em Ricci flow} ) first appeared in the
mathematics literature in a paper of H-D Cao \cite{Cao} and has been
discussed independently in the physics context by Adams, Polchinski
and Silverstein \cite{APS} (the exact solution also appears in
\cite{GHMS}). This example has led to the use of RG flow to
qualitatively study tachyon condensation.

These results strongly suggest a link between (in)stability of a
vacuum geometry with respect to a suitable RG flow and its off-shell
(in)stability \footnote{In fact, the orbifold flow example and ideas
in \cite{HMT} suggest that (in)stability with respect to RG flow may
give an approximate indication of on-shell (in)stability as well.}.
Motivated by this, we propose an investigation of off-shell
stability of vacua in closed string theory by studying their
stability with respect to various RG flows. The study of off-shell
stability of vacua using string field theory techniques is a hard
problem. The study of stability of a geometry with respect to an RG
flow, while not an easy problem, is mathematically well-posed in
many cases. It may offer insights in a target space, rather than
world-sheet picture, of the types of instabilities of string vacua.
The world-sheet RG flows of closed string theory are {\em geometric
flows}, the simplest of these (truncated to first order in
$\alpha'$) being the Ricci flow. The Ricci flow is a subject of
active research in mathematics and has been used to prove the
Poincar\'e conjecture (for papers related to this, see
\cite{perelman}, \cite{rf}). There are also stability results for
various geometries with respect to Ricci flow (we discuss some of
these in section II). Thus, there is an excellent opportunity to
generalize these techniques (as well as techniques from relativity)
to address stability of geometries with respect to world-sheet RG
flows \footnote{It has even been suggested in \cite{HW} that the
thermodynamic instability of a black hole may be related to its
instability under Ricci flow.}. The stability analysis for various
geometries in relativity can be adapted to study stability with
respect to geometric flows as well. A stability analysis of special
solutions (such as the hairpin brane) to RG flows of Dirichlet sigma
models has been done recently in \cite{Bakas}.

One of the most well-studied vacuum geometries in string theory is
of the form $AdS_{p} \times S^q$ obtained as a stationary point of a
supergravity action (for specific $p$ and $q$)
--- an investigation of its stability with respect to RG flow would therefore
be interesting \footnote{Classical stability of $AdS_p \times M^q$
geometries appearing as Freund-Rubin compactifications has been
studied in \cite{dewolfe}.}. Ideally the stability of $AdS_{p}
\times S^q$ ought to be studied with respect to an RG flow for which
it is a fixed point. However, we recall that the curvature of
$AdS_{p} \times S^q$ is usually sourced by the RR field. Although
there are world-sheet actions with a component that is the sigma
model on $AdS_{p} \times S^q$ (see section II for a review), there
exists no computation of a $\beta$ function with an RR field.
$AdS_{3}$ is a fixed point of RG flow with a NS-NS $B$ field
\cite{hw}. However this construction does not generalize to $AdS$
space in higher dimensions.

In this paper, as a first step, we address the stability of $n$
dimensional (Euclidean) AdS space (hyperbolic space $\mathbf H^n $)
with respect to RG flow \footnote{For a complete discussion of
classical stability of $AdS$ spacetime in the context of relativity,
see \cite{ishiwald}.}. The RG flow we choose is the Ricci flow.
$AdS$ space expands uniformly under Ricci flow. One can study if
perturbations of $AdS$ space decay, and if the perturbed geometry
approaches $AdS$ space under Ricci flow (up to overall scale). This
notion of stability is called {\em geometric stability} of $AdS$
space with respect to Ricci flow. We prove geometric stability of
$AdS_{n}$ with respect to Ricci flow for perturbations obeying
specific asymptotic behaviour. The techniques we use are motivated
by recent work of Andersson and Moncrief \cite{am} for compact
hyperbolic space in the context of relativity
--- these can be generalized to study noncompact spaces. The
world-sheet sigma model with $AdS$ target space is not a CFT,
however a stability result with respect to Ricci flow should be
indicative of stability results with respect to RG flows with
additional fields (for which $AdS_{n}$ may be a fixed point). The
$AdS_n$ sigma model was also studied by Friess and Gubser \cite{fg},
who obtained the $\beta$ function in a particular large $n$
expansion. We comment in the last section of the paper on some
curious connections between our result and theirs. The techniques we
use to investigate stability with respect to Ricci flow can be
generalized to RG flow with a $B$ field. Throughout the paper, we
assume that the truncation of the RG flow to first order (in
$\alpha'$) is a valid approximation. When the curvature of the
geometry is of order $1/\alpha'$, the sigma model perturbation
theory breaks down. In this case, other effects, such as those
caused by winding tachyons become important. In the context of
asymptotically $AdS$ spaces, this is discussed in \cite{hs}.

 The plan of the paper is as follows: Section II contains a
summary, for physicists, of a set of techniques to investigate
stability of fixed points of RG flows. One of the techniques is used
in the mathematics of Ricci flow to investigate stability of compact
manifolds. We outline another technique which can be applied to
investigate stability of noncompact manifolds and then describe the
study of geometric stability of $AdS_n $ using these techniques.
Finally we discuss the motivations for such a study from string
theory. This section summarizes the results and the steps involved
in proving them, for readers who may want to skip the mathematical
details.

Section III and the Appendices contain the main results, proofs, a
discussion on boundary terms, asymptotic conditions on the
perturbation, and the associated mathematical details. Section IV is
a discussion of future projects and some connections with results in
\cite{fg}.

\section{Stability techniques }
 \setcounter{equation}{0}
The Ricci flow arises in physics as the simplest lowest-order (in
square of string length $\alpha'$) RG flow of the world-sheet sigma
model for closed strings. In this context, the Ricci flow is the
flow of the metric of the target space with respect to the RG flow
parameter $t$. Both in physics and mathematics, we are interested in
a flow of geometries (i.e., of metrics mod diffeomorphisms).
Therefore all flows related to each other by $t$ dependent
diffeomorphisms generated by a vector field $V$ are equivalent; we
write the generic flow in this class --- the {\em Ricci-de Turck}
flow as:
\begin{eqnarray}
\frac{\partial g_{ij}}{\partial t} = - \alpha' (R_{ij} + \nabla_{i}
V_j + \nabla_{j} V_i ) \label{2.1}
\end{eqnarray}

 Ricci flow in two and three dimensions on compact manifolds is
now well-understood --- with curvature conditions on an initial
geometry, much is known about the limiting geometry under Ricci flow
\cite{chow, topping}. This is not the case for solutions to Ricci
flow in higher dimensions, or on noncompact manifolds. Stability
results for geometries under small perturbations are therefore
useful in such cases. An innovative stability technique due to Cao,
Hamilton and Ilmanen \cite{CHI} applies monotonicity formulae under
Ricci flow derived by Perelman \cite{perelman} (the $\lambda$
functional and shrinker entropy). The $\lambda$ functional, for
instance, is defined as follows: First, consider a compact manifold
$M$ with metric $g$. Define the functional
\begin{eqnarray}
\mathcal{F}(g, f) := \int_M e^{-f} (|\nabla f|^2 + R)~dV \label{2.2}
\end{eqnarray}
Then, the $\lambda$ functional is defined by
\begin{eqnarray}
\lambda(g) := \inf \{ \mathcal{F}(g, f) : f \in C_{c}^{\infty} (M),
\int e^{-f} ~dV = 1  \}, \label{2.3}
\end{eqnarray}
and has some nice properties. Under Ricci flow, $\lambda(g)$ is
constant only on fixed points of geometry \footnote{These
geometries, which are fixed points of Ricci-de Turck flow, are
called {\em steady Ricci solitons}. It can be shown that the only
steady Ricci solitons on {\em compact} manifolds are Ricci
flat.\cite{bour}} and is monotonically increasing otherwise. As
shown in \cite{CHI}, these properties can then be used to
investigate linear stability of a fixed point $g$. Consider a
perturbation at $t=0$, $g^{p}(0) = g(0) + h(0).$ Let $\lambda(g) =
c$. This is constant along Ricci flow. Let $\lambda(g^{p}(0))$ be
the value of the $\lambda$ functional for the perturbed geometry at
$t=0$. If $\lambda(g^{p}(0)) - c > 0$, then clearly
$\lambda(g^{p}(t))$ can only increase for $t > 0$. So the perturbed
geometry will not approach the fixed point. The sign of
$\lambda(g^{p}(0)) - c $ can therefore be used to determine
stability of the fixed point. This is evaluated in a linearized
approximation where $h(0)$ is assumed small, and it gives a notion
of linear stability when the sign is negative. A similar analysis
works for the shrinker entropy, which is constant on {\em shrinking
Ricci solitons} (i.e., geometries that shrink uniformly under
Ricci-de Turck flow), and is monotonically increasing otherwise. The
shrinker entropy can be used to determine `geometric stability' of
shrinking Ricci solitons. Consider a perturbation of a shrinking
Ricci soliton. If the perturbed geometry approaches that of the
shrinking Ricci soliton under Ricci-de Turck flow (up to overall
scale), we say the soliton is geometrically stable. Using the
shrinker entropy, it is shown in \cite{CHI} that $\mathbf{S}^{n}$ is
geometrically stable. For results on stability analysis of other
geometries, see \cite{CHI}, \cite{GIK}, \cite{sesum},\cite{DWW}. A
discussion of stability of Einstein metrics under Ricci flow appears
in \cite{rye}. A stability result for flat space under Ricci flow is
derived in \cite{SSS}. The linear stability of {\em compact}
hyperbolic space under (a normalized) Ricci flow is discussed in the
appendix to \cite{KY}. For a generalization of the stability
analysis of \cite{CHI} to RG flow with a $B$ field, see \cite{OSW1}.

It is interesting to note that in the context of relativity, a paper
by Fischer and Moncrief that appeared earlier (than \cite{perelman})
uses closely related stability techniques to analyze stability of
certain Einstein metrics \cite {fm}. The authors of \cite{fm}
construct a monotonic quantity in time (the reduced Hamiltonian)
that is constant only on a class of self-similar solutions to the
Einstein equation. The reduced Hamiltonian can be used to obtain
linear stability results as in the case of Ricci flow.

 An `expander entropy' similar
to the shrinker entropy, for expanding Ricci solitons has been
constructed in \cite{Ni}. The $\lambda$ functional and these
entropies are defined on compact manifolds, and cannot be
generalized easily to noncompact manifolds. It is possible that such
entropies could be defined with some curvature conditions on the
geometry under the flow --- for an example on asymptotically flat
spaces with negative curvature, see \cite{OSW}. However, if we wish
to examine stability of noncompact manifolds in general, we need
other techniques. In the context of relativity, stability of compact
hyperbolic space has been examined by Andersson and Moncrief
\cite{am} --- the techniques used can be generalized to study
noncompact manifolds. They can also be adapted to study stability
with respect to geometric flows like the Ricci flow. We will outline
the techniques we use in this paper, which are motivated by those in
\cite{am}. Start with a fixed point geometry $g$ under a geometric
flow. Consider a perturbed geometry $g^{p}(t) = g(t) + h(t)$. We
assume $h(t)$ is small so that we can work with a linearized flow
equation for $h$. We then define a particular Sobolev norm of the
perturbation $h$ given in terms of integrals of the perturbation and
its spatial derivatives as defined in Eq.(\ref{22}). One of the main
tasks is to bound the Sobolev norm under the linearized flow and
prove that the Sobolev norm goes to zero as $t \rightarrow \infty$.
This can be done by defining some integrals we call `energies' ---
which are given by the squared $L^2$ norms of the perturbation and
its derivatives with respect to $t$ (see Eq. (\ref{12})). We first
show the energies decay under the flow and then use this to prove
that the Sobolev norm decays under the flow. In many cases, there
are results generically termed {\em Sobolev inequalities} or {\em
Sobolev embeddings} that then imply that the perturbation and its
derivatives decay {\em pointwise} everywhere on the manifold. This
gives a linear stability result for the fixed point $g$ with respect
to the geometric flow. On noncompact manifolds, we may need to
prescribe asymptotic behaviour for the perturbation and its
derivatives to deal with boundary terms that arise while studying
the evolution of the Sobolev norm. We prescribe the asymptotic
behaviour at initial $t$ and check that this asymptotic form of the
perturbation is preserved by the flow equation.

We can apply this technique to prove geometric (linear) stability of
a geometry such as Euclidean $AdS_n$ ($\mathbf{H}^n $) with respect
to Ricci flow. We first study its linear stability with respect to
the flow Eq. (\ref{8}), chosen so that $\mathbf{H}^n $ is a fixed
point. The solutions to Eq. (\ref{8}) are related to those of Ricci
flow with a rescaled metric and flow parameter. The linear stability
of $\mathbf{H}^n$ with respect to Eq. (\ref{2}) implies geometric
linear stability with respect to Ricci flow. The linear stability of
$\mathbf{H}^n $ with respect to the flow given by Eq. (\ref{2}) is
proved as we have outlined, by showing that the Sobolev norm of the
perturbation is bounded under the flow.

As discussed in the introduction, the main motivation is a better
understanding of the off-shell stability of $AdS_n$ through an
investigation of its stability with respect to a suitable RG flow.
There are various sigma model descriptions of $AdS$ space. $AdS_3$
is a fixed point of the RG flow of the sigma model with metric and
NS-NS field (with the $H$ field proportional to the volume form)
\cite{hw}. $AdS_3 \times S^3 \times T^4$ is obtained as a
near-horizon limit of the sigma model with a $B$ field describing
the NS-5 brane (this is an exact solution in heterotic string
theory) \cite{ct}. The sigma model with $AdS_3 \times S^3$ target
space has been studied in \cite{GKS}, for example.  The sigma model
with $AdS_n$ target space has been analyzed in \cite{fg}. There is a
sigma model description of $AdS_5 \times S^5$ as well, by Metsaev
and Tseytlin \cite{mt}. They propose a world-sheet action that
accounts for the RR field as well; and whose bosonic part is the
usual sigma model on $AdS_5 \times S^5$. However, there is no
computation of a $\beta$ function
--- they provide arguments for why the action
should be conformal. Since an RG flow with an RR field is not known,
one cannot study the stability of $AdS_5 \times S^5$ using sigma
model RG flows with this geometry as fixed point. However, one can
still choose an RG flow such as Ricci flow, with respect to which
this geometry evolves, and analyze geometric stability. This should
be indicative of the stability under metric perturbations with
respect to other RG flows as well.

We conclude this section by remarking that it may also be possible
to generalize (classical) stability calculations in relativity and
supergravity to study stability of a wider class of geometries under
RG flows. For example, Gibbons and Hartnoll \cite{gh} discuss
classical stability of a class of geometries (that include some
higher dimensional black holes) under certain types of
perturbations. They show that the stability under these
perturbations is related to the spectrum of the Lichnerowicz
laplacian on a lower dimensional manifold, the `base' manifold.
However, unlike this classical stability analysis, we cannot gauge
away the trace of a perturbation when studying stability under RG
flows.

There has been extensive work on studying classical perturbations of
$AdS$ spacetime itself. For example, in \cite{ishiwald}, it is shown
that the equations governing scalar, vector and tensor perturbations
of $AdS$ spacetime all reduce to the action of a positive,
symmetric, elliptic differential operator on a space of
square-integrable functions. These equations can be solved with
appropriate boundary conditions (the perturbations vanish at the
conformal boundary). This then implies that $AdS$ spacetime is
classically linearly stable. Further, given an assumption on the
solutions to the linearized Einstein equation for $AdS$ spacetime -
namely that (up to infinitesimal diffeomorphisms) they are uniquely
determined by their series expansion at the conformal boundary - it
is possible to obtain certain global uniqueness results for the
$AdS$ spacetime \cite{anderson}. For a proof that the above
assumption holds for conformally compact Riemannian-Einstein
metrics, see \cite{andersonherzlich}.

In the stability problem we discuss in this paper, the geometric PDE
governing the evolution of perturbations of (Euclidean) $AdS$ space
is the linearized Ricci flow, not the linearized Einstein equation.
Nevertheless, it would be desirable to solve the linear stability
problem of $AdS$ space under Ricci flow completely, with suitable
boundary conditions, rather than the analysis based on the Sobolev
norms, and the asymptotic behaviour we have imposed. It is also
possible that a local existence result can be obtained for the
linearized flow for initial data in some appropriate Sobolev space.
We hope to return to this question in future work.

\section{ Linearized analysis of stability of hyperbolic space }
 \setcounter{equation}{0}
The conventions we follow in the paper for the curvature tensors are
those of Wald \cite{wald}.

We investigate the stability of $n$ dimensional hyperbolic space
$\mathbf{H}^n$ with respect to the Ricci flow
\begin{eqnarray}
\frac{\partial \tilde g_{ij}}{\partial \tilde t} = - \alpha' \tilde R_{ij}.
\label{1}
\end{eqnarray}
 $\tilde
t$ is the RG flow parameter.\\


$\mathbf{H}^n$ is not a fixed point of the Ricci flow. It is
convenient to first study its stability with respect to a flow with
$\mathbf{H}^n$ as fixed point, such that its solutions are related
to those of Ricci flow. We consider the flow
\begin{eqnarray}
\frac{\partial g_{ij}}{\partial t} = - \alpha' [ R_{ij} + c g_{ij} ]
\label{2}
\end{eqnarray}
whose solutions are related to those of (\ref{1}) by
\begin{eqnarray}
\tilde t = \frac{1}{\alpha' c}~~ e^{\alpha' c t}, \nonumber \\
\tilde g_{ij} = e^{\alpha' c t} g_{ij}.
\label{3}
\end{eqnarray}

\noindent Recall that the Riemann tensor of $\mathbf{H}^n$ is
\begin{equation}
R_{ijkl} = -\frac{1}{a} [ g_{ik} g_{jl} - g_{il} g_{jk}],
\label{4}
\end{equation}
where $a > 0$ is a constant.
Therefore,
\begin{eqnarray}
R_{ij} = - \frac{(n-1)}{a} g_{ij}, \nonumber \\
R = - \frac{n(n-1)}{a}.
\label{5}
\end{eqnarray}
$\mathbf{H}^n$ is a fixed point of Eq.(\ref{2}) with $c =
\frac{(n-1)}{a}$.\\


We now wish to consider perturbations of the metric $g$ of
$\mathbf{H}^n$. The perturbed metric $g_{ij}^{p} = g_{ij} + h_{ij}$.
We will study the evolution of the perturbation under the flow
Eq.(\ref{2}) (more precisely, a flow related to it by
diffeomorphisms) in a linearized approximation. We will then show in
the rest of the section that the perturbation decays under the flow.
$g_{ij}^{p}(t)$ is related to a solution of Ricci flow (in the
linearized approximation) by the rescalings Eq.(\ref{3}). The
solution to Ricci flow is $\tilde g_{ij}(\tilde t) = e^{\alpha' c t}
(g_{ij} + h_{ij}(t))$; and if we show that $h_{ij}(t) \rightarrow 0$
as $t \rightarrow \infty$, then the perturbed geometry approaches
that of $\mathbf{H}^n$ under Ricci flow in this approximation, up to
the overall scale $e^{\alpha' c  t}$. This is our precise notion of
geometric linear stability.

Under the flow Eq.(\ref{2}), in a linearized approximation, the
perturbation $h_{ij}$ obeys
\begin{eqnarray}
\frac{\partial h_{ij}}{\partial t} = \frac{\alpha'}{2}
[ (\Delta_{L} h)_{ij} + \nabla_{i}\nabla_{j} H - \nabla_{i}(\nabla^{k}h_{kj})
- \nabla_{j}(\nabla^{k}h_{ki}) - 2 c h_{ij} ].
\label{6}
\end{eqnarray}
Here and in what follows, all covariant derivatives are taken with
respect to the background metric $g$. $H = g^{ij} h_{ij}$ is the
trace of the perturbation.
\begin{eqnarray}
(\Delta_{L} h)_{ij} = \Delta h_{ij} + 2 R_{kilj} h^{kl} - R_{i}^{k} h_{jk}
- R_{j}^{k} h_{ik}
\label{7}
\end{eqnarray}
is the Lichnerowicz laplacian acting on symmetric 2-tensors (all
curvature tensors being those of the background metric).

A $t$ dependent diffeomorphism generated by vector field $X$
modifies the flow (\ref{2}) for the metric $g^{p}$ to
\begin{eqnarray}
\frac{\partial g_{ij}^{(p)}}{\partial t} = - \alpha' [ R_{ij}^{(p)}
+ c g_{ij}^{(p)} + \nabla_{i}^{(p)} X_j + \nabla_{j}^{(p)} X_i ]
\label{8}
\end{eqnarray}
where (\ref{8}) is now written in the new coordinates \footnote{We
can relate solutions to this modified flow with solutions to an
appropriate Ricci -de Turck flow of the form Eq.(\ref{2.1}) by
rescalings as in Eq.(\ref{3}).}.

$X$ can be chosen so that when we now linearize (\ref{8}), we can
eliminate the `divergence' terms that appear in the linearized flow
Eq.(\ref{6}), $\frac{\alpha'}{2} \left ( \nabla_{i}\nabla_{j}H -
\nabla_{i}(\nabla^{k}h_{kj}) - \nabla_{j}(\nabla^{k}h_{ki}) \right
)$, leading to the simplified flow (in the new
coordinates),\footnote{This is part of the `de-Turck trick' in Ricci
flow used to prove short-time existence of solutions. See page 79,
\cite{chow} for details.}
\begin{eqnarray}
\frac{\partial h_{ij}}{\partial t} = \frac{\alpha'}{2} [ (\Delta_{L}
h)_{ij} - 2 c h_{ij} ]. \label{9}
\end{eqnarray}
Denote
\begin{eqnarray}
\frac{\partial h_{ij}}{\partial t} = (\Lambda h)_{ij} \label{10}
\end{eqnarray} where
\begin{eqnarray}
(\Lambda~ h)_{ij} = \frac{\alpha'}{2} \left [ (\Delta_{L} h)_{ij}
- 2 c h_{ij} \right ] \nonumber \\
= \frac{\alpha'}{2} \left [ \Delta h_{ij} - \frac{2}{a} H g_{ij} +
\frac{2}{a} h_{ij} \right ]. \label{11}
\end{eqnarray}
\\
Now, it can be easily checked that when the background metric is
that of $\mathbf{H}^n$, the operator $\Lambda$ acting on the
`divergence' part of the perturbation preserves this form. Such
terms can be removed by a suitable $t$ dependent diffeomorphism as
we did before. In the rest of the paper, we therefore only consider
divergence-free (i.e., transverse) perturbations. Such a
perturbation $h_{ij}$ satisfies $\nabla^{j} h_{ij} = 0$.


Let us now consider the following general `energy integral'
$E^{(K)}$, defined by
\begin{eqnarray}
E^{(K)} = \int_M \left |(\Lambda^{(K)} h)_{ij} \right |^2 ~ dV ,
\label{12}
\end{eqnarray}
where the notation $( \Lambda^{(2)} h )_{ij} = (\Lambda \Lambda
h)_{ij}$, for example (denote $(\Lambda^{(0)}h)_{ij} = h_{ij}$).
Further, we define the `cross term'
\begin{eqnarray}
E^{(K, K+1)} = \int_M (\Lambda^{(K+1)} h)^{ij} (\Lambda^{(K)}
h)_{ij} ~dV \label{12a}\end{eqnarray}

We will bound $E^{(K)}$ and $E^{(K,K+1)}$ under the flow. In
order to do this, we need the following lemmas:\\

\noindent {\bf Proposition 3.1:}
 If $h_{ij}$ is a transverse perturbation, then $(\Lambda^{(K)}
h)_{ij}$ is transverse on $\mathbf H^n $.\\

\noindent {\bf Proof:} Consider a transverse tensor $T_{ij}$ (i.e.,
$\nabla^j T_{ij} = 0$). Let $Tr~T$ denote $g^{ij}T_{ij}$. Then
\begin{eqnarray}
\nabla^j (\Lambda T)_{ij} = \frac{\alpha'}{2} \nabla^{j} \left [
\Delta T_{ij} - \frac{2}{a}(Tr~T)g_{ij} + \frac{2}{a} T_{ij} \right
] \label{13}
\end{eqnarray}
\begin{eqnarray}
\nabla^j (\Delta T)_{ij} &=& \Delta (\nabla^j T_{ij}) + g^{pk}
g^{js} \left [ \nabla_p \left ( R_{ski}^{~~~m} T_{mj} +
R_{skj}^{~~~m}T_{im} \right ) + \right. \nonumber \\
&& \left. R_{spk}^{~~~m} \nabla_m T_{ij} + R_{spi}^{~~~m} \nabla_{k}
T_{mj} + R_{spj}^{~~~m} \nabla_k T_{im}
\right ] \nonumber \\
&=&\frac{2}{a} \nabla_{i} (Tr~T). \label{14} \end{eqnarray}
Therefore, it follows from (\ref{13}) and (\ref{14}) that $\nabla^j
(\Lambda T)_{ij} = 0$. Setting $T_{ij} = h_{ij}$, $(\Lambda h)_{ij}$
is transverse, and it follows that $(\Lambda^{(K)} h)_{ij}$ is
transverse for all positive integers $K$. \qed \\

\noindent {\bf Lemma 3.2:} For a transverse symmetric tensor
$T_{ij}$, we have the following inequality on $\mathbf H^n$:
\begin{gather}
\int_M (\Delta T_{ij})T^{ij} ~dV \leq  \frac{1}{a} \int_M (Tr~T)^2
~dV
-  \frac{n}{a} \int_M |T_{ij}|^2 ~dV \nonumber \\
+ \int_{\partial M} \left [ (\nabla_k T_{ij}) - (\nabla_j T_{ik})
\right ] n^k T^{ij}~dA \label{15}
\end{gather}\\

\noindent {\bf Proof:} Consider the manifestly nonnegative quantity
$\epsilon$ :
\begin{gather}
\epsilon = \int_M \left | \nabla_k T_{ij} - \nabla_j T_{ik} \right |^2 ~dV \nonumber \\
= 2 \int_M \left |\nabla_k T_{ij} \right |^2 ~dV - 2 \int_M
(\nabla_k T_{ij} )(\nabla^{j} T^{ik}) ~dV \label{16}
\end{gather}
Now,
\begin{gather}
\int_M (\nabla_k T_{ij} )(\nabla^{j} T^{ik}) =
\int_{\partial M} (\nabla_k T_{ij})T^{ik} n^j ~dA \nonumber \\
- \int_M R_{jkim}T^{mj} T^{ik} ~dV - \int_M R_{k}^{m} T_{im} T^{ik}
~dV, \label{17}
\end{gather}
where we integrated by parts, commuted two covariant derivatives,
and used the fact that $T$ is a transverse tensor.

On $\mathbf{H}^n$, we therefore have
\begin{gather}
\int_M (\nabla_k T_{ij} )(\nabla^{j} T^{ik}) = \int_{\partial M}
(\nabla_k T_{ij})T^{ik} n^j ~dA -\frac{1}{a} \int_M (Tr~T)^2 ~dV +
\frac{n}{a} \int_M |T_{ik}|^2 ~dV \label{18}
\end{gather}
Similarly,
\begin{gather}
\int_M \left |\nabla_k T_{ij} \right |^2 ~dV = \int_{\partial M}
(\nabla_k T_{ij})T^{ij} n^k ~dA - \int_M (\Delta T_{ij})T^{ij} ~dV
\label{19}
\end{gather}
Therefore, from (\ref{17}---\ref{19}),
\begin{gather}
\epsilon = 2 \int_{\partial M} (\nabla_k T_{ij})T^{ij} n^k ~dA
- 2 \int_M (\Delta T_{ij})T^{ij} ~dV \nonumber \\
- 2 \int_{\partial M} (\nabla_k T_{ij})T^{ik} n^j dA +  \frac{2}{a}
\int_M (Tr~T)^2 ~dV - \frac{2n}{a} \int_M |T_{ik}|^2 ~dV \label{20}
\end{gather}
Since $\epsilon \geq 0$, the inequality in the lemma follows. \qed \\

We are now ready to prove the following bound for the energies: \\

\noindent {\bf Theorem 3.3:}
On $\mathbf H^n$,\\
\begin{eqnarray}
 \frac{dE^{(K)}}{dt} \leq - \alpha' \frac{(n-2)}{a} E^{(K)} + \alpha'
\int_{\partial M} \left [ (\nabla_k T_{ij}) - (\nabla_j T_{ik})
\right ] n^k T^{ij}~dA \label{21}
\end{eqnarray}
where, in the inequality above (and in the proof), $T_{ij} =
(\Lambda^{(K)} h)_{ij}$.\\

\noindent {\bf Proof:}
\begin{eqnarray}
\frac{dE^{(K)}}{dt} &=& 2 \int_{M} (\Lambda^{(K)} h)^{ij}
(\Lambda^{(K+1)} h)_{ij} ~dV \nonumber \\
&=& \alpha' \int_M T^{ij} \left [ \Delta T_{ij} - \frac{2}{a} (Tr~T)
g_{ij} + \frac{2}{a} T_{ij} \right ] ~dV \nonumber \\
& \leq & - \alpha' \frac{(n-2)}{a} \int_M T^{ij} T_{ij} ~dV -
\alpha' \frac{1}{a} \int_M (Tr~T)^2 ~dV +  \nonumber \\ && \alpha'
\int_{\partial M} \left [ (\nabla_k T_{ij}) - (\nabla_j T_{ik})
\right ] n^k T^{ij}~dA \nonumber \\
&\leq& - \alpha' \frac{(n-2)}{a} E^{(K)} + \nonumber \\ && \alpha'
\int_{\partial M} \left [ (\nabla_k T_{ij}) - (\nabla_j T_{ik})
\right ] n^k T^{ij}~dA, \nonumber \\
\end{eqnarray}
where the inequality in the last two steps follows from Proposition
3.1 and Lemma 3.2 . \qed \\

We note that for $n=2$, we only need to consider the flow of the
trace of the perturbation. We can construct `energies' as above from
the $t$ derivatives of the trace and bound them under the flow. This
is discussed in Appendix D. In fact, we can prove that the trace
decays pointwise on the manifold, and this is enough to prove
geometric linear stability of $\mathbf{H}^2$.

We now define a particular {\em Sobolev norm} of the perturbation
(see Appendix B for a discussion of Sobolev norms), denoted by
$\parallel h
\parallel_{k,2}$, by the following equation:
\begin{eqnarray}
\left ( \parallel h \parallel_{k,2} \right )^2 &=& \int_M |h_{ij}|^2 ~dV + \int_M
|\nabla_{p_1} h_{ij}|^2 ~dV + ....\nonumber \\&&...+ \int_M
|\nabla_{p_1}....\nabla_{p_k} h_{ij} |^2 ~dV \label{22}
\end{eqnarray}
In the above expression, $|h_{ij}|^2$, for example, is the square of
the (pointwise) tensor norm of the perturbation, i.e.,
$h^{ij}h_{ij}$, and indices are raised with respect to the
background metric $g$. We assume that this norm is defined for our
perturbation at initial $t$.

From Lemma A1.1 in the Appendix, the square of the above Sobolev
norm can be written in a form more useful for our purposes (for $k$
an even positive integer) and $\{ A_{i} \}$ for $i=1,..,k$
constants:
\begin{eqnarray}
\left ( \parallel h \parallel_{k,2} \right )^2 &=& A_0 \int_M |h_{ij}|^2 ~dV + A_1
\int_M h^{ij} \Delta h_{ij} ~dV + A_2 \int_M |\Delta h_{ij}|^2 ~dV....\nonumber \\
&&....+ A_k \int_M | \Delta^{\frac{k}{2}} h_{ij}|^{2} ~dV \nonumber \\
&&+(similar~sum~over~integrals~of~traces~of~above~tensors) \nonumber
\\ &&+ ~b.t . \label{23}
\end{eqnarray}
$b.t$ is an abbreviation for boundary terms.
When $k$ is odd, we
have, similarly,
\begin{eqnarray}
\left ( \parallel h \parallel_{k,2} \right )^2 &=& A_0 \int_M |h_{ij}|^2 ~dV + A_1
\int_M h^{ij} \Delta h_{ij} ~dV +......\nonumber \\
&&....+ A_k \int_M (\Delta^{\frac{(k-1)}{2}} h^{ij})(\Delta^{\frac{(k+1)}{2}}h_{ij}) ~dV \nonumber \\
&&+(similar~sum~over~integrals~of~traces~of~above~tensors) \nonumber
\\ &&+~ b.t. \label{24}
\end{eqnarray}

The boundary terms that appear in the discussion above as well as
everywhere else in the paper are discussed in Appendix C (we will
impose appropriate asymptotic behaviour on the perturbations later
so that the boundary terms vanish). By Corollary A1.3, we can study
the evolution of $\left ( \parallel h
\parallel_{k,2} \right )^2 $ under the flow by studying the
evolution of the integral $\int_M (\Lambda^{P} h)^{ij} (\Lambda^{Q}
h)_{ij} ~dV$ for arbitrary integers $P,Q \geq 0$. This task is made
easier by the following
Proposition; \\

\noindent {\bf Proposition 3.4 :} Let $P,Q \geq 0$ be two integers
such that $(P-Q) = R \geq 0$. Let $T_{ij} = (\Lambda^{Q} h)_{ij}$.
$\int_M (\Lambda^{P} h)^{ij} (\Lambda^{Q} h)_{ij} ~dV = \int_M
T^{ij} (\Lambda^{R} T)_{ij} ~dV $.\\

(i) If $R$ is odd, then
\begin{eqnarray}
\int_M T^{ij} (\Lambda^{R} T)_{ij} ~dV = \int_M
(\Lambda^{\frac{(R-1)}{2}} T)^{ij} (\Lambda^{\frac{(R+1)}{2}}
T)^{ij} ~dV +~ b.t . \label{25}
\end{eqnarray}\\
(ii) If $R$ is even, then
\begin{eqnarray}
\int_M T^{ij} (\Lambda^{R} T)_{ij} ~dV = \int_M
(\Lambda^{\frac{R}{2}} T)^{ij} (\Lambda^{\frac{R}{2}} T)^{ij} ~dV +~
b.t . \label{26}
\end{eqnarray}
\\
\noindent {\bf Proof:} From Eq.(\ref{11}),
\begin{eqnarray}
(\Lambda^{R} T)_{ij} = \frac{\alpha'}{2} \left [ \Delta
(\Lambda^{(R-1)} T)_{ij} - \frac{2}{a} Tr~(\Lambda^{(R-1)} T)
~g_{ij} + \frac{2}{a} (\Lambda^{(R-1)} T)_{ij} \right ] \label{27}
\end{eqnarray}
\begin{eqnarray}
\int_M T^{ij} \Delta (\Lambda^{(R-1)} T)_{ij} ~dV = \int_M \Delta
T^{ij}(\Lambda^{(R-1)} T)_{ij} ~dV +~b.t , \label{28} \end{eqnarray}
by integrating by parts. Therefore it follows from Eq.(\ref{27}) and
Eq.(\ref{28}) that
\begin{eqnarray}
\int_M T^{ij} (\Lambda^{R} T)_{ij} ~dV = \int_M (\Lambda
T)^{ij}\Lambda^{(R-1)} T)_{ij} ~dV +~b.t , \label{29}
\end{eqnarray}
and that we can iterate this procedure until we get the righthand
sides of Eq.(\ref{25}) and Eq.(\ref{26}). \qed \\

\noindent {\bf Corollary 3.5 :} Thus, modulo boundary terms, from
Corollary A1.3 and Proposition 3.4, it follows that $\left (
\parallel h \parallel_{k,2} \right )^2 $ can be written as a sum over integrals
which are precisely the energies $E^{(K)}$ and `cross terms'
$E^{(K,K+1)}$ that we defined earlier. \qed \\


We now discuss the asymptotic behaviour that is required of the
perturbation for the various boundary terms (listed in Appendix C)
to vanish. We would like to impose such asymptotic behaviour at some
$t = t_0$ and then argue that it is preserved by the flow at all $t
> t_0$. To do this, it is convenient to start with the Poincar\'e
ball metric for $\mathbf H^n$
--- i.e., $\mathbf H^n $ is the interior of the n-dimensional unit ball with metric
$g = e/\rho^2$ where $e$ is the Euclidean metric. Choosing Cartesian
coordinates $\{ x^i \}$,
\begin{eqnarray}
ds^2 = \frac{1}{\rho^2} \sum_{i=1}^{n} (dx^{i})^2 .
\label{30}
\end{eqnarray}
$\rho = (1-r^2)/(2 \sqrt{a})$. $r = \sqrt{ \sum_{i=1}^{n} (x^{i})^2 }$.
The volume element $dV = \rho^{-n} dx^n $. The geodesic distance $R(x)$ from
the origin to $x$ is
\begin{eqnarray}
R = \sqrt{a} \ln \left ( \frac{1+r}{1-r} \right )
\label{31}
\end{eqnarray}

Then, asymptotically as $ r \rightarrow 1$, $\rho \approx
(1-r)/(\sqrt{a})$, and $\exp(-R/(\sqrt{a})) \approx \frac{1}{2}~
(1-r) $. We restrict ourselves in the rest of the paper to
perturbations for which, as $ r \rightarrow 1$, $h_{jk} \sim
O((1-r)^{\beta})$ implies $\partial_{i}h_{jk} \sim O((1-r)^{\beta -
1})$. Then, a study of the boundary terms listed in Appendix C
reveals that the asymptotic behaviour required for the boundary
terms to vanish is $h_{jk} \sim O((1-r)^{\frac{n-4}{2} +
\epsilon})$, $\epsilon
> 0$. We can impose such a condition at some $t= t_0$. We can
then check whether the evolution operator $\Lambda$ preserves this
asymptotic behaviour. Since $h_{jk}(t_{0}) \sim
O((1-r)^{\frac{n-4}{2} + \epsilon})$ as $r \rightarrow 1$, it is
easily seen (because the Poincar\'e ball metric is conformal to the
Euclidean metric) that $(\Lambda h)_{jk} \sim O((1-r)^{\frac{n-4}{2}
+ \epsilon})$ asymptotically. Therefore, studying Eq.(\ref{11}) as
$r \rightarrow 1$, it is clear that $h_{ij}(t)$ obeys this
asymptotic behaviour for $t > t_{0}$. Due to the (simple) nature of
the evolution operator $\Lambda$, it is probable that a more
rigorous local existence result can be obtained by starting with an
initial condition $h \in H_{s}^{2}(T_{2}M, g)$\footnote{See Appendix
B for the definition of the Sobolev space $ H_{s}^{2}(T_{2}M, g)$.}
and adapting some of the results in \cite{Taylor} to the flow
equation Eq.(\ref{11}).

We observe that for $n \ge 4$, the asymptotic behaviour required is
a `fall-off', and corresponds to the perturbation falling off
exponentially as a function of the geodesic distance from the origin
$R(r)$ as $r \rightarrow 1$. The asymptotic conditions that
characterize asymptotically anti-de Sitter spaces also correspond to
exponential fall-offs in the geodesic distance from the origin.
However, these are obtained in different coordinates, in for e.g.,
\cite{ht} and a comparison of the two sets of asymptotic behaviour
is not straightforward, due to the fact that the two boundaries have
different topologies.
\\


\noindent {\bf Corollary 3.6 :} For solutions $h_{ij}(t)$ of
Eq.(\ref{11}) that obey the asymptotic behaviour
$h_{jk}(t) \sim O((1-r)^{\frac{n-4}{2} + \epsilon})$ as $r \rightarrow 1$,\\

(i) It follows from Theorem 3.3 that
\begin{eqnarray}
E^{(K)}(t) \leq E^{(K)}(t_0)~e^{ -D(t-t_{0})} ,
\label{32}
\end{eqnarray}
where $D = \alpha'~ (n-2)/a $.
\\

(ii) The cross term $E^{(K, K+1)}(t)$ obeys the inequality
\begin{eqnarray}
E^{(K, K+1)} (t) \leq \sqrt{E^{(K)}(t)E^{(K+1)}(t)}
 \leq \sqrt{E^{(K)}(t_0 )E^{(K+1)}(t_0 )} ~e^{ -D(t-t_{0})}~~~~
\label{33} \end{eqnarray} This follows from the Cauchy-Schwarz
inequality for vector spaces with inner product $<.,.>$ and norm
$\parallel . \parallel$, $<T,W> ~ \leq~ \parallel T
\parallel  \parallel W \parallel$, when we consider the $L^2 $ norm of
tensors.
\\

(iii) A consequence of (i), (ii) and Corollary 3.5 is that the
Sobolev norm $\parallel h \parallel_{k,2} $ is bounded with respect
to the flow Eq.(\ref{11}), and $\parallel h \parallel_{k,2}(t)
\rightarrow 0$ as $t \rightarrow \infty $. \qed \\

We now use certain Sobolev-type inequalities derived in the context of
 $\mathbf H^n $ by L. Andersson \cite{la}. {\em Sobolev inequalities} are
generically inequalities between various norms (a useful reference
for Sobolev inequalities for domains in $\mathbf R^n$ is
\cite{Adams} ). We are particularly interested in a Sobolev
inequality that bounds the $C^k$ norm of the perturbation from above
by a constant times its Sobolev norm (see Appendix B for the
definition of the $C^k$ norm). The Sobolev inequalities in \cite{la}
for $\mathbf H^n $ are obtained by first showing that the Sobolev
norm on  $\mathbf H^n $ --- i.e., on the unit ball with metric $g =
e/\rho^{2}$ is equivalent to a {\em weighted} Sobolev norm on the
unit ball with metric $e$ with suitable weights functions which are
powers of $\rho$ (see Appendix B for definition of a weighted
Sobolev norm). The result most useful for us from \cite{la} is the
following inequality for $s
> n/2$ ($s$ is a nonnegative integer):
\begin{eqnarray}
\parallel h \parallel_{C^k} ~ \leq C \parallel h \parallel_{k+s,~ 2} .
\label{34}
\end{eqnarray}

$C$ is a constant. We have already seen that $ \parallel h
\parallel_{k+s,~ 2}$ (the Sobolev norm of the perturbation of
$\mathbf H^n $ ) is bounded under the flow Eq. (\ref{11}) and decays
exponentially as $t \rightarrow \infty$. Therefore, choosing $s$
large enough, inequality Eq.(\ref{34}) implies that the perturbation
and its derivatives up to order $k$ are bounded {\em pointwise} on
$M$, and decay as $t \rightarrow \infty$. Thus $\mathbf H^n$ is
linearly stable with respect to the flow Eq.(\ref{11}) for
perturbations obeying the asymptotic conditions we have prescribed.
Further, we know from Eq.(\ref{3}) that such perturbations of
$\mathbf H^n$ also decay under Ricci flow and the geometry
approaches that of $\mathbf H^n$ up to an overall scale --- so this
implies geometric linear stability for $\mathbf H^n$ with respect to
Ricci flow.

\section{Discussion}
 \setcounter{equation}{0}

From the discussion so far, we conclude that $AdS_n$ and
$\mathbf{S}^n$ are geometrically (linearly) stable with respect to
Ricci flow. The stability of $AdS_p \times S^q $ with respect to
Ricci flow still remains to be investigated --- this problem is not
straightforward since hyperbolic space expands under Ricci flow
while the sphere contracts. However, we at least know that this
geometry is stable when the (metric) perturbation is restricted to
either $AdS_p$ or $S^q$. A stability result with respect to Ricci
flow is indicative of a stability result with respect to other RG
flows, when restricted to only metric perturbations. In the absence
of an RG flow involving the RR field, the only other computation we
can now attempt is to check stability of $AdS_3$ with respect to RG
flow with a $B$-field. It should be possible to do this computation
by generalizing the energies defined in section III and using
similar techniques. If $AdS_3$ is stable with respect to this RG
flow (restricting to only metric perturbations) as we expect, this
would offer yet another reason why $AdS$ space is interesting in
string theory. It would also be indicative of off-shell stability
(with respect to some processes at least) in string field theory.

What does this stability result for $AdS_n$ imply for its quotients?
In three dimensions, one of the quotients of $AdS$ space is a black
hole geometry --- the BTZ black hole \cite{btz}. The stability
result for $AdS_n$ required a careful analysis of boundary terms and
an imposition of asymptotic behaviour on the perturbations. Since
the BTZ black hole is obtained from $AdS_{3}$ by global
identifications, this analysis will be different for the BTZ black
hole, and thus its stability remains an open question.

Is it possible to generalize this result on linear stability to a
notion of {\em nonlinear} stability? We use a linearized flow and
therefore require the perturbation $h(t)$ to be small. The linear
stability results of \cite{CHI} (on compact manifolds) with respect
to Ricci flow, on the other hand, only require the perturbation to
be small at some initial $t$. The diffeomorphism-invariant
functionals and entropies then allow a study of the perturbed
geometry with respect to the (full nonlinear) Ricci flow. We would
like to generalize the linear stability result in this paper to a
result where we only require the perturbation to be small at initial
$t$ --- this is a notion of nonlinear stability, since the perturbed
geometry is evolving under the nonlinear flow. We therefore need to
either extend or find an analogue of Perelman's entropies for
noncompact manifolds
--- this may be possible with some curvature conditions on the
geometry under the flow. In the context of relativity, Andersson and
Moncrief prove nonlinear stability of compact hyperbolic space in
this sense \cite{am}
--- hopefully these techniques can be generalized to study nonlinear
stability of $\mathbf{H}^n$ with respect to the geometric flows
considered in this paper.

Finally, we comment on a curious (possibly coincidental) connection
to a study of the $AdS_n$ sigma model by Friess and Gubser
\cite{fg}. These authors observe that the effective expansion
parameter in the sigma model is $\kappa = - \frac{\alpha'(n-1)}{a}$.
They compute the $\beta$ function of the sigma model in the large
$n$ limit but to all orders in $\alpha'$ (such that
$\frac{\alpha'(n-1)}{a}$ is kept fixed). The authors find a
nontrivial $AdS_n$ fixed point at second order in this expansion for
a particular value of $\kappa$. Curiously, the parameter $\kappa$ in
\cite{fg} is {\em exactly the same} as the decay constant
characterizing the (exponential) rate of decay of the trace of the
perturbation $H = g^{ij}h_{ij}$ in our analysis (see Appendix D for
this derivation). We do not know yet of a logical argument
connecting these two computations, since ours is a target space
analysis, whereas the computation in \cite{fg} is done entirely on
the world-sheet.
\section{Acknowledgements}
I am indebted to Vincent Moncrief for introducing me to the
stability techniques used in this paper and for various enjoyable
discussions at every stage of this project. I thank Lars Andersson
for useful comments and for bringing \cite{la} to my attention. I
also thank Eric Woolgar for a discussion on asymptotic analysis and
for comments on a draft of the paper. I wish to acknowledge other
comments on this work and related topics by Ioannis Bakas, Huai-Dong
Cao, Viqar Husain, K Narayan, Todd Oliynyk and Bala Sathiapalan.
This work was initiated during a visit to Albert Einstein Institute,
Golm and I thank the institute for hospitality. This work is
supported by funds from the Natural Sciences and Engineering
Research Council of Canada.

\section{Appendix A: Useful results}
 \setcounter{equation}{0}

Here we prove some Lemmas required for section 3 of the paper.\\

\noindent {\bf Lemma A1.1 :} For some choice of constants $\{ B_{i}
\}$, $i= 1,...,k$, and $k$ a positive integer, on $\mathbf H^n $,\\

(i) When $k$ is even,
\begin{eqnarray}
\int_M |\nabla_{p_1}....\nabla_{p_k} h_{ij} |^2 ~dV &=& B_0 \int_M
|h_{ij}|^2 ~dV + B_1
\int_M h^{ij} \Delta h_{ij} ~dV + B_2 \int_M |\Delta h_{ij}|^2 ~dV\nonumber \\
&&....+ B_k \int_M | \Delta^{\frac{k}{2}} h_{ij}|^2 ~dV \nonumber \\
&&+(sum~over~integrals~of~traces~of~above~tensors) \nonumber
\\&&+ b.t . \label{a1-1}
\end{eqnarray}\\

(ii) When $k$ is odd,
\begin{eqnarray}
\int_M |\nabla_{p_1}....\nabla_{p_k} h_{ij} |^2 ~dV &=& B_0 \int_M
|h_{ij}|^2 ~dV + B_1
\int_M h^{ij} \Delta h_{ij} ~dV +......\nonumber \\
&&....+ B_k \int_M (\Delta^{\frac{(k-1)}{2}} h^{ij})(\Delta^{\frac{(k+1)}{2}}h_{ij}) ~dV \nonumber \\
&&+(sum~over~integrals~of~traces~of~above~tensors)
\nonumber \\ && + b.t . \label{a1-2} \end{eqnarray}\\

\noindent {\bf Proof:} We prove this result by induction. First note
that
\begin{eqnarray}
\int_M |\nabla_{p} h_{ij} |^2 ~dV &=& - \int_M h^{ij} \Delta h_{ij}
~dV \nonumber \\&& + \int_{\partial M} h^{ij} (\nabla_p h_{ij}) n^p
~dA. \label{a1-3}
\end{eqnarray}

Similarly,
\begin{eqnarray}
\int_M |\nabla_{p}\nabla_{s} h_{ij} |^2 ~dV &=& \int_{\partial M}
(\nabla_{p}\nabla_{s} h^{ij})(\nabla^s h_{ij})n^{p} ~dA
- \int_M \nabla^s h^{ij} \Delta (\nabla_s h_{ij}) ~dV,\nonumber \\
&=& \int_{\partial M} (\nabla_{p}\nabla_{s} h^{ij})(\nabla^s
h_{ij})n^{p} ~dA + \int_M \nabla^s h^{ij} \nabla_s (\Delta h_{ij})
~dV \nonumber \\
&& - \frac{(n-1)}{a} \int_M |\nabla_p h_{ij}|^2 ~dV - \frac{8}{a}
\int_M (\nabla^s h^{ij})(\nabla_{i} h_{sj}) ~dV.~~~~~~~~~~~
\label{a1-4}
\end{eqnarray}

Now,
\begin{eqnarray}
- \frac{8}{a} \int_M (\nabla_s h_{ij})(\nabla^{i} h^{sj}) ~dV &=& -
\frac{8}{a} \int_{\partial M} (\nabla_s h_{ij}) h^{sj} n^{i} ~dA +
\frac{8}{a} \int_M h^{sj}(\nabla^i \nabla_s h_{ij}) ~dV \nonumber \\
&=& - \frac{8}{a} \int_{\partial M} (\nabla_s h_{ij}) h^{sj} n^{i}
~dA - \frac{8(n+1)}{a^2} \int_M |h_{sj}|^2 ~dV \nonumber \\&&+
\frac{16}{a^2} \int_M H^2 ~dV. \label{a1-5}
\end{eqnarray}
In the above expression, we commuted two covariant derivatives in
the final step and used the expressions for the curvature tensors of
the background. Further,
\begin{eqnarray}
\int_M \nabla^s h^{ij} \nabla_s (\Delta h_{ij}) ~dV &=&
\int_{\partial M} \nabla_s h^{ij} (\Delta h_{ij}) n^s ~dA - \int_M
|\Delta h_{ij}|^2 ~dV. ~~~~~~~~~~~\label{a1-6} \end{eqnarray}

Using (\ref{a1-5}) and (\ref{a1-6}) in (\ref{a1-4}), we see that the
statement of Lemma A1.1 is indeed satisfied.\\

Now assume the induction hypothesis, i.e., that both parts of the
statement of Lemma A1.1 are true for all $\int_M
|\nabla_{p_1}....\nabla_{p_m} h_{ij} |^2 ~dV$, and $m \leq k$. Let
us first assume $k$ is even. Then, consider $m=(k+1)$. First, let
$A_{p_{2}...p_{k}ij} = \nabla_{p_{2}}...\nabla_{p_{k}} h_{ij}$.
Then,
$\nabla_{p_1}....\nabla_{p_k} h_{ij} =
\nabla_{p_{1}}A_{p_{2}...p_{k}ij}$.

\begin{eqnarray}
&&\int_M |\nabla_{q} \nabla_{p_1}....\nabla_{p_k} h_{ij} |^2 ~dV =
\int_M \nabla^{q} \nabla^{p_1}A^{p_{2}...ij}\nabla_{q}
\nabla_{p_1}A_{p_{2}...ij} ~dV , \nonumber \\
&=& - \int_M \nabla^{p_1}A^{p_{2}...ij} \Delta
\nabla_{p_1}A_{p_{2}...ij} ~dV + ~b.t.~~~~~~~~~~~~~~~ \nonumber \\
&=& - \int_M \nabla^{p_1} A^{p_{2}...j} \nabla_{p_1} \Delta
A_{p_{2}...j} ~dV + \frac{(n-1)}{a} \int_M
|\nabla_{p_1}A_{p_{2}...ij} |^2 ~dV \nonumber \\
&+& \frac{2}{a} \sum_{l=2}^{k} \int_M \nabla^{p_1}
A^{p_{2}..p_{l}..j} \nabla_{p_{l}} A_{p_{2}..p_{1}..j}~dV -
\frac{2}{a} \sum_{l=2}^{k} \int_M \nabla^{p_1} A^{p_{2}..j}
\nabla^{m} A_{p_{2}..m..j}~ g_{p_{1} p_{l}} ~dV \nonumber \\
&+& \frac{4}{a} \int_M \nabla^{p_1}A^{p_{2}..ij} \nabla_{i}
A_{p_{2}..p_{1}j} ~dV - \frac{4}{a} \int_M
\nabla^{p_1}A^{p_{2}...ij} \nabla^{r} A_{p_{2}...rj}~g_{p_{1}i} ~dV
\nonumber \\ && + b.t~~~~~~~~~~~~ \label{a1-7}\end{eqnarray}

Now we look at the structure of each term in the last line of
(\ref{a1-7}). Consider
\begin{eqnarray}
I_1 = \int_M \nabla^{p_1} A^{p_{2}...j} \nabla_{p_1} \Delta
A_{p_{2}...j} ~dV = - \int_M \Delta A^{p_{2}...j}\Delta
A_{p_{2}...j} ~dV + b.t~~~~~
\label{a1-8}\end{eqnarray} \\
Writing $A_{p_{2}...p_{k}ij} = \nabla_{p_{2}}...\nabla_{p_{k}}
h_{ij}$ we can again commute $\Delta$ and $\nabla_{p_{2}}$ in the
integrand of $I_{1}$. Further, by a sequence of such steps, we can
write this term as a sum of $\int_M \Delta^{\frac{k}{2}}h^{ij}
\Delta^{\frac{(k+2)}{2}}h_{ij} ~dV $ and similar terms with lower
powers of $\Delta$ to which the induction hypothesis applies. This
is because every step where the covariant derivatives are commuted
produces Riemann curvature terms. In $\mathbf H^n $, these are
simply proportional to products of metric components and produce
traces of pairs of derivative operators or of the perturbation. This
applies to other terms in the last line of (\ref{a1-7}) as well. For
the term
\begin{eqnarray}
I_2 =  \frac{(n-1)}{a} \int_M |\nabla_{p_1}A_{p_{2}...ij} |^2 ~dV
\label{a1-9}
\end{eqnarray}
the induction hypothesis applies directly. For
\begin{eqnarray}
I_3 &=& \int_M \nabla^{p_1} A^{p_{2}..p_{l}..j} \nabla_{p_{l}}
A_{p_{2}...p_{1}..j} ~dV, \nonumber \\
I_4 &=& \int_M \nabla^{p_1} A^{p_{2}...j} \nabla^{m}
A_{p_{2}...m..j}~ g_{p_{1} p_{l}} ~dV \label{a1-10}
\end{eqnarray}
again one can commute derivatives and the argument in the previous
paragraph applies.

The terms
\begin{eqnarray}
I_5 &=& \frac{4}{a} \int_M \nabla^{p_1}A^{p_{2}...ij} \nabla_{i}
A_{p_{2}...p_{1}j} ~dV = - \frac{4}{a} \int_M \nabla_{i}
\nabla^{p_1}A^{p_{2}...ij} A_{p_{2}...p_{1}j} ~dV +
b.t~,\nonumber \\
I_6 &=& \int_M \nabla_{i} A^{p_{2}...ij} \nabla^{r} A_{p_{2}...rj}
~dV \label{a1-11}
\end{eqnarray}
are somewhat different. Since $A_{p_{2}...p_{k}ij} =
\nabla_{p_{2}}...\nabla_{p_{k}} h_{ij}$, we can commute derivatives,
till we get a term with $\nabla_{i}h^{ij}$ which is zero. Other
terms are proportional to the Riemann tensor, as before. Thus it is
easy to see that the statement of Lemma A1.1 is true, by these
obvious steps. The same can be done if $k$ is odd. \qed \\

\noindent {\bf Lemma A1.2 :} For $p$ a positive integer, and some
constants $c_{K}$, $K = 0,...,p$,
\begin{eqnarray}
\Delta^{p} h_{ij} = \sum_{K=0}^{p} c_{K} (\Lambda^{(K)} h)_{ij} +
\sum_{K=0}^{p} \tilde c_{K} Tr~(\Lambda^{(K)} h) g_{ij} .
\label{a1-12}
\end{eqnarray}
We have used the notation $(\Lambda^{0} h)_{ij} = h_{ij}$ and as
usual, $Tr~(\Lambda^{(K)} h) = (\Lambda^{(K)} h)_{ij} g^{ij}$.\\

\noindent {\bf Proof:} The case $p = 1$ is obviously true from
Eq.(\ref{11}). We can write
\begin{eqnarray}
\Delta h_{ij} = \frac{2}{\alpha'} (\Lambda h)_{ij} + \frac{2}{a} H
g_{ij} - \frac{2}{a} h_{ij}. \label{a1-13} \end{eqnarray}

Now assume the statement of the lemma is true for $\Delta^{p}$ for
some $p$. Then it is easy to see that the statement of the lemma is
true for $\Delta^{p+1} h_{ij}$. From Eq.(\ref{11}),
\begin{eqnarray}
\Delta^{p+1} h_{ij} = \frac{2}{\alpha'} (\Lambda (\Delta^{p}
h))_{ij} + \frac{2}{a} (Tr~\Delta^{p} h) g_{ij} - \frac{2}{a}
\Delta^{p} h_{ij}, \label{a1-14}
\end{eqnarray} and one can then
substitute for $\Delta^{p} h_{ij}$ from Eq.(\ref{a1-12}). \qed \\

\noindent{\bf Corollary A1.3 :} It therefore follows that an
integral of the form $\int_M | \Delta^{\frac{k}{2}} h_{ij}|^2 ~dV$
or $\int_M (\Delta^{\frac{(k-1)}{2}}
h^{ij})(\Delta^{\frac{(k+1)}{2}}h_{ij}) ~dV$ can be written as a sum
of integrals of the form $\int_M (\Lambda^{P} h)^{ij} (\Lambda^{Q}
h)_{ij} ~dV$ for integers $P,Q \geq 0$ \qed.

\section{Appendix B: Sobolev and weighted Sobolev spaces}
 \setcounter{equation}{0}
Consider a manifold $M$ with Riemannian metric $g$. We define
Sobolev spaces of sections of tensor bundles over $M$. Let us first
recall the definition of various norms. For a section $v$ of
$T_{s}^{r}M$ we can define the pointwise tensor norm by using the
metric. Denote this by $|v|_{g}$. Define, for $ 1 \leq p < \infty$,
\begin{eqnarray}
\parallel v \parallel_{L^p} = \left ( \int_M  ( |v|_{g} )^{p} ~dV_{g} \right )^{\frac{1}{p}}
\label{a2-1} \end{eqnarray}
to be the $L^p$ norm of $u$ when $ \int_M  ( |v|_{g} )^{p} ~dV_{g} < \infty$.
Define $\parallel v \parallel_{\infty} = $inf$ \{K: |v(x)|\leq K ~$almost everywhere on$~ M \}$.
This is also called the essential supremum of $|v|$. Now let $u$ be a section
of $T_{s}^{r}M$ of class $C^k$.
Let $\alpha = (\alpha_{1},....\alpha_{j})$ be the $j$-tuple of
non-negative integers $\alpha_{i}$. We call $\alpha$ a multi-index.
$\nabla^{\alpha} = \nabla_{1}^{\alpha_{1}}...\nabla_{j}^{\alpha_{j}}$ is a
differential operator of order $|\alpha| = \sum_{i=1}^{j} \alpha_{i}$.
The $C^k$ norm of u is defined as
\begin{eqnarray}
\parallel u \parallel_{C^k} = \max_{0 \leq |\alpha| \leq k}~\sup_{x \in M}
|\nabla^{\alpha} u(x)|,
\label{a2-2} \end{eqnarray}
Define the Sobolev, or $H_{k}^{p}$ norm as
\begin{eqnarray}
\parallel u \parallel_{k,p} = \left ( \sum_{ 0 \leq |\alpha| \leq k} \int_M  (| \nabla^{\alpha} u|_{g} )^{p} ~dV_{g} \right )^{\frac{1}{p}} ,
\label{a2-3} \end{eqnarray}
if  $ 1 \leq p < \infty$. Also define $\parallel u \parallel_{k,\infty} =
$max$_{0\leq |\alpha| \leq k} \parallel \nabla^{\alpha}u \parallel_{\infty}$.

The Sobolev space $H_{k}^{p}\left (T_{s}^{r}M, g) \right )$ is the
completion of $\{ u ~$a section of$~ T_{s}^{r}M ~$of class$~C^k  : \parallel u \parallel_{k,p} < \infty \}$
with respect to the norm $\parallel.\parallel_{k,p}$. We shall abbreviate this in the
paper as $H_{k}^{p}$ wherever it is obvious that we are dealing with
tensors on $M$.
$H_{k}^{2}\left (T_{s}^{r}M, g \right )$ is a Hilbert space.

Define now the
weighted Sobolev norm of $u$ with weight functions $w_{\alpha}$ as
\begin{eqnarray}
\parallel u \parallel_{k,p, w} = \left ( \sum_{ 0 \leq |\alpha| \leq k} \int_M  (| \nabla^{\alpha} u|_{g} )^{p} ~w_{\alpha}~dV_{g} \right )^{\frac{1}{p}} ,
\label{a2-4} \end{eqnarray}
if  $ 1 \leq p < \infty$.
The weighted Sobolev spaces are defined using this norm, similar to Sobolev
spaces.

\section{ Appendix C: Boundary terms}
\setcounter{equation}{0}

We list the various types of boundary terms that arise in the
analysis of the stability of $\mathbf{H}^n$.

Let $T_{ij} = (\Lambda^{(K)}h)_{ij}$.

The boundary terms associated with the proof of Lemma 3.2 are of the
form
\begin{eqnarray}
B_{1} = \int_{\partial M} \left[ (\nabla_k T_{ij}) - (\nabla_j
T_{ik}) \right ] n^k T^{ij} ~dA. \label{a3-1} \end{eqnarray}

The boundary terms associated with the proof of Proposition 3.4 are
of the form
\begin{eqnarray}
B_{2} = \int_{\partial M} T^{ij} \nabla_p (\Lambda^{(Q)}h)_{ij} n^p
~dA.\label{a3-2} \end{eqnarray}
\begin{eqnarray}
B_3 = \int_{\partial M} \nabla_p T^{ij} (\Lambda^{(Q)}h)_{ij} n^p
~dA. \label{a3-3} \end{eqnarray}

The boundary terms arising from the proof of Lemma A1.1 are of the
form
\begin{eqnarray}
B_4 = \int_{\partial M} \nabla^{p_1}...(\Delta)^{R}..\nabla^{p_k}
h^{ij} \nabla_s \nabla_{p_1}..(\Delta)^{Q}..\nabla_{p_k} h_{ij}
n^{s} ~dA. \label{a3-4} \end{eqnarray}
\begin{eqnarray}
B_5 = \int_{\partial M} \nabla^{p_1}...(\Delta)^{R}..\nabla^{p_k}
h^{sj} \nabla_s \nabla_{p_1}..(\Delta)^{Q}..\nabla_{p_k} h_{ij}
n^{i} ~dA. \label{a3-5} \end{eqnarray}

$(\Delta)^{2} A_{i_1...i_k} = \Delta \Delta A_{i_1...i_k}$ according
to our notation. The asymptotic behaviour required to eliminate
these boundary terms is given in section III.

\section{Appendix D: Flow of the trace of the perturbation}
\setcounter{equation}{0}

Taking the trace of Eq.(\ref{9}) with respect to the background
metric, we get the flow of the trace of the perturbation,
\begin{eqnarray}
\frac{\partial H}{\partial t} = \frac{\alpha'}{2} ~[ \Delta H - 2 c
H] = L~H.\label{ad-1}
\end{eqnarray}

We now define the energy integral,
\begin{eqnarray}
E_{H}^{(K)} = \int_M (L^{(K)}H)^2 ~dV, \label{ad-2}
\end{eqnarray}
where $L^{(2)}H = L L H$. We can study the flow of $E_{H}^{(K)}$.
\begin{eqnarray}
\frac{\partial E_{H}^{(K)}}{\partial t} &=& 2 \int_M  L^{(K)}H \frac{\partial L^{(K)}H}{\partial t} ~dV, \nonumber \\
&=& \alpha' \int_M  L^{(K)}H [ \Delta - 2 c]L^{(K)}H ~dV, \nonumber \\
&=& -2 c \alpha' E_{H}^{(K)} - \alpha' \int_M |\nabla_{i} (L^{(K)}H)
|^2 ~dV + \nonumber \\ && \alpha' \int_{\partial M} (L^{(K)}H)
\nabla_{i}(L^{(K)}H) n^i ~dA,
\nonumber \\
& \leq & -2 \frac{\alpha'(n-1)}{a} E_{H}^{(K)} + \alpha'
\int_{\partial M} (L^{(K)}H) \nabla_{i}(L^{(K)}H) n^i
~dA.~~~~~~~~~~~~~
 \label{ad-3}
\end{eqnarray}

The asymptotic behaviour discussed in section III guarantees that
the boundary term in (\ref{ad-3}) vanishes. Then we can reproduce
the rest of the analysis --- we can construct the Sobolev norm of
$H$, argue using the energy bound (\ref{ad-3}) that this norm decays
under the flow of the trace (\ref{ad-1}), and that $H$ decays
pointwise on $M$. Clearly the decay of $H$ must be faster than or
equal to $C e^{-(\alpha'(n-1)/a)t}$, where $C$ is some constant.
Note that the rate of exponential decay governed by $-\alpha'(n-1)/a
$ is the same as the constant $\kappa$ in \cite{fg}.


\begin{thebibliography}{100}
\bibitem{HMT} M Headrick, S Minwalla, T Takayanagi, Class Quant Grav
21 (2004) S1539.
\bibitem{APS} A Adams, J Polchinski, E Silverstein, JHEP 0110 (2001)
029.
\bibitem{OZ} Y Okawa, B Zwiebach, JHEP 0403 (2004) 056.
\bibitem{Cao} H-D Cao, J Diff Geom 45 (1997) 257.
\bibitem{GHMS} M Gutperle, M Headrick, S Minwalla, V Schomerus, JHEP
0301 (2003) 073.
\bibitem{perelman} G Perelman, arXiv: math/0211159, math/0303109,
math/0307245.
\bibitem{rf} H-D Cao, X-P Zhu, Asian Journal of Mathematics 10
(2006) 165; J Morgan, G Tian, {\em Ricci flow and the Poincare
Conjecture}, Clay Math Monographs Volume 3, CMI, 2007.
\bibitem{HW} M Headrick, T Wiseman, Class.Quant.Grav. 23 (2006)
6683. See also G. Holzegel, T. Schmelzer, C. Warnick, Class Quant
Grav 24 (2007) 6201.
\bibitem{Bakas} I Bakas, C Sourdis, arXiv:0704.3985.
\bibitem{dewolfe} O DeWolfe, DZ Freedman, SS Gubser, GT Horowitz, I
Mitra, Phys Rev D65 (2002) 064033; SS Gubser, I Mitra, JHEP 0207
(2002) 044; T Shiromizu, D Ida, H Ochiai, T Torii, Phys Rev D64
(2001) 084025.
\bibitem{ishiwald} A Ishibashi, RM Wald, Class Quant Grav 21 (2004)
2981.
\bibitem{hw} G Horowitz, D Welch, Phys Rev Lett 71 (1993) 328.
\bibitem{am} L Andersson, V Moncrief, {\em Einstein spaces as
attractors for the Einstein flow}, unpublished.
\bibitem{fg} J Friess, S Gubser, Nucl Phys B750 (2006) 111.
\bibitem{hs} GT Horowitz, E Silverstein, Phys.Rev. D73 (2006)
064016.
\bibitem{chow} B Chow, {\em The Ricci flow: an introduction},
Mathematical Surveys and Monographs, Volume 110, AMS, 2004.
\bibitem{topping} P. Topping, {\em Lectures on the Ricci flow},
London Mathematical Society lecture note series 325, Cambridge
University Press, 2006.
\bibitem{CHI} H-D Cao, R Hamilton, T Ilmanen, arXiv: math/0404165.
\bibitem{bour} J-P Bourguignon in {\em Global differential geometry
and global analysis}, Lectures in Mathematics 838, ed. D Ferus,
Springer, 1981.
\bibitem{GIK} C Guenther, J Isenberg, D Knopf, Comm Anal Geom 10
(2002) no.4, 741; Int. Math. Res. Not. (2006), Article ID 96253,
doi: 10.1155/IMRN/2006/96253.
\bibitem{sesum} N Sesum, arXiv: math/0410062.
\bibitem{DWW} X Dai, X Wang, G Wei, arXiv: math/0504527.
\bibitem{rye} R Ye, Trans Amer Math Soc 338 (1993) 871.
\bibitem{SSS} O Schnuerer, F Schulze, M Simon, Comm Anal Geom 16
(2008) 127.
\bibitem{KY} D Knopf, A Young, Proc. Amer. Math. Soc., to appear.
\bibitem{OSW1} T Oliynyk, V Suneeta, E Woolgar, Nucl Phys B739 (2006)
441.
\bibitem{fm} AE Fischer, V Moncrief, Class Quant Grav 19 (2002)
5557.
\bibitem{Ni} M Feldman, T Ilmanen, L Ni, arXiv: math/0405036.
\bibitem{OSW} T Oliynyk, V Suneeta, E Woolgar, Phys Lett B610 (2005)
115.
\bibitem{ct} M Cvetic, AA Tseytlin, Phys Lett B366 (1996) 95.
\bibitem{GKS} A Giveon, D Kutasov, N Seiberg, Adv Theor Math Phys 2
(1998) 733.
\bibitem{mt} RR Metsaev, AA Tseytlin, Nucl Phys B533 (1998) 109.
\bibitem{gh} GW Gibbons, SA Hartnoll, Phys Rev D66 (2002) 064024;
see also GW Gibbons, SA Hartnoll, CN Pope, Phys Rev D67 (2003)
084024.
\bibitem{anderson} MT Anderson, Class Quant Grav 23 (2006) 6935.
\bibitem{andersonherzlich} MT Anderson, M Herzlich, J Geom Phys, 58, (2008),
179.
\bibitem{wald} RM Wald, {\em General Relativity}, The University of
Chicago Press, 1984.
\bibitem{Taylor} ME Taylor, {\em Partial Differential Equations
vol. 3: Nonlinear equations}, Springer, New York, 1996.
\bibitem{ht} M Henneaux, C Teitelboim, Comm Math Phys 98, no. 3
(1985 ) 391.
\bibitem{la} L Andersson, Indiana U. Math. J. 42, no.4 (1993) 1359.
\bibitem{Adams} RA Adams, JJF Fournier, {\em Sobolev spaces, 2nd
Ed.}, vol. 140, Pure and Applied Mathematics Series, Academic Press,
2003.
\bibitem{btz} M Banados, C Teitelboim, J Zanelli, Phys Rev Lett 69
(1992) 1849; M Banados, M Henneaux, C Teitelboim, J Zanelli, Phys
Rev D48 (1993) 1506.

\end{thebibliography}
\end{document}